\documentclass[aps,prl,reprint,superscriptaddress]{revtex4-1}

\usepackage[T1]{fontenc}
\usepackage{graphicx}
\usepackage{amssymb}

\begin{document}

\newcommand{\didv}{$\mathrm{d}I/\mathrm{d}V$}
\newcommand{\tbpc}{$\mathrm{TbPc_2}$}
\newcommand{\tb}{$\mathrm{Tb^{3+}}$}

\title{Nuclear and electronic resonance spectroscopy of single molecules by radio-frequency scanning tunnelling microscopy}

\author{Stefan M\"{u}llegger}
\email[]{stefan.muellegger@jku.at}
\author{Stefano Tebi} 
\affiliation{Institute of Semiconductor and Solid State Physics, Johannes Kepler University Linz, 4040 Linz, Austria.}
\author{Amal K. Das}
\altaffiliation{On sabbatical leave from the Department of Physics and Meteorology, Indian Institute of Technology Kharagpur-721302, India.}
\affiliation{Institute of Semiconductor and Solid State Physics, Johannes Kepler University Linz, 4040 Linz, Austria.}
\author{Wolfgang Sch\"{o}fberger}
\affiliation{Institute of Organic Chemistry, Johannes Kepler University Linz, 4040 Linz, Austria.}
\author{Felix Faschinger}
\affiliation{Institute of Organic Chemistry, Johannes Kepler University Linz, 4040 Linz, Austria.}
\author{Reinhold Koch}
\affiliation{Institute of Semiconductor and Solid State Physics, Johannes Kepler University Linz, 4040 Linz, Austria.}

\maketitle

The ongoing miniaturization in nanoscience and \mbox{-technology} challenges the sensitivity and selectivity of experimental analysis methods to the ultimate level of single atoms and molecules. 
A promising new approach, addressed here, focuses on the combination of two well-established complementary techniques that have proven to be very successful in their own fields:  
(i) low-temperature scanning tunneling microscopy (STM), offering high spatial resolution for imaging and spectroscopy together with the capability of manipulating single atoms and molecules in a well-controlled manner \cite{Chen2008};  
(ii) radio-frequency (rf) magnetic resonance techniques, providing paramount analytical power based on a high energy resolution combined with the versatility of being sensitive to a great variety of different properties of matter \cite{Abragam1970,Slichter1996,Levitt2001}. 
Here, we demonstrate the successful resonant excitation and detection of nuclear and electronic magnetic transitions of a single quantum spin in a single molecule by rf tunneling of electrons applied through the tip of a modified STM instrument operated at 5~K. 
The presented \mbox{rf-STM} approach allows the unrivalled spectroscopic investigation of electronic hyperfine levels in single molecules with simultaneous sub-molecular spatial resolution. 
The achieved single-spin sensitivity represents a ten orders of magnitude improvement compared to existing methods of magnetic resonance -- offering, atom-by-atom, unprecedented analytical power and spin control with impact to physics, chemistry, biology, medicine, nanoscience and -technology. 

While classical electron- and nuclear magnetic resonance methods utilize electromagnetic fields to excite electron and nuclear spin transitions, electric currents in a magnetic system can efficiently excite its magnetic moments, as well. 
For instance, dc electron tunneling has been demonstrated to excite magnons \cite{Balashov2006}, flip the spin of a single atom \cite{Heinrich2004,Loth2010}, or switch on and off molecular spins \cite{Komeda2011} by inelastic processes. 
Rf currents through nanoscale magnetic bars have been shown to excite ferromagnetic resonance \cite{Fang2011}. 
Compared to these earlier approaches, the resonant spin excitation in single molecules, described in the present work, is based on simultaneous dc tunneling through molecular electronic levels while superposing an additional rf tunneling current. 
Tuning the rf component in resonance with different spin transitions is found to affect the molecular electron tunneling conductance. 
Thereby, conductance measurements during rf excitation enable a new type of ``magnetic resonance'' spectroscopy at low magnetic fields of only a few mT and excitation frequencies ranging presently between $\approx10$~MHz and 4~GHz. 

\begin{figure*}
\includegraphics[width=16cm]{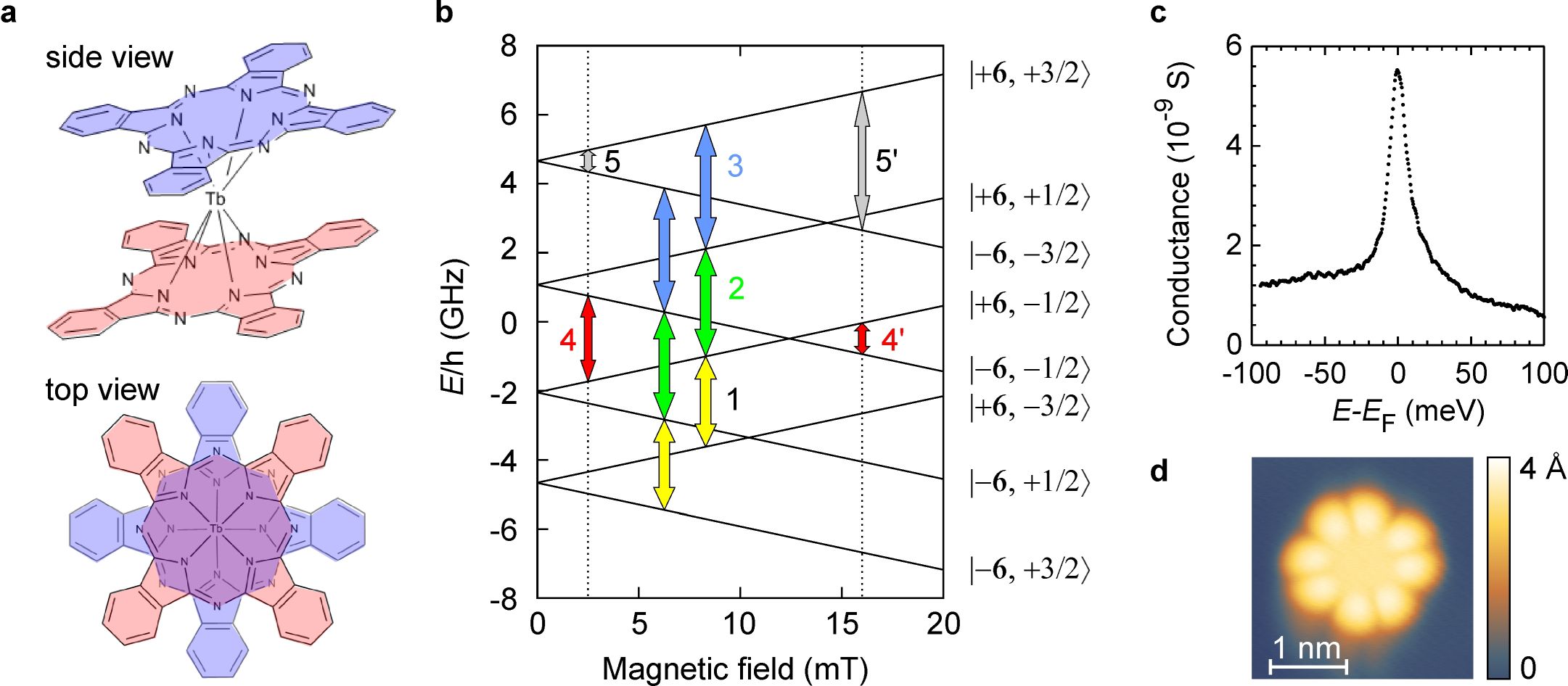}
\caption{\label{fig:tb} Neutral ``terbium double decker'' molecule, [\tbpc]$^0$. (a) Molecular structure; the two phthalocyanine ligands are colored red and blue for clarity. (b) Zeeman diagram of the lowest $J_\mathrm{z}=\pm6$ substates of the $J=\pm6$ electronic ground-state of Tb$^{3+}$ in a neutral [\tbpc]$^0$ molecule; nuclear hyperfine interaction with the $I=3/2$ Tb nucleus splits each electronic level $J_z$ into four hyperfine levels labeled as $\left|J_z,I_z\right\rangle$; yellow, green and blue arrows mark purely nuclear hyperfine transitions with $\Delta J_z=0$ and $\Delta I_z=\pm1$, labeled 1--3; red arrows mark a purely electronic hyperfine transition with $\Delta J_z=\pm12$ and $\Delta I_z=0$ at two different field values, labeled 4 and 4', respectively; grey arrows mark a selected mixed electronic-nuclear transition at different field values, labeled 5 and 5', respectively; dotted lines indicate two field values realized in our experiments. (c) Kondo signature in tunneling conductance spectrum characteristic of [\tbpc]$^0$ adsorbed on Au(111). (d) STM topographic image of a single [\tbpc]$^0$ molecule on Au(111) at $+0.9$~V.} 
\end{figure*}

As model system for rf-STM based spin resonance, we have chosen the well-known ``terbium double decker'' molecule bis(phthalocyaninato)terbium(III) (\tbpc, Fig.~\ref{fig:tb}a), consisting of a Tb$^{3+}$ ion sandwiched between two organic phthalocyanine ligands \cite{Ishikawa2003a,Ishikawa2003b}.  
The Tb$^{3+}$ ion has an electronic configuration of [Xe]$4f^8$ resulting in a total spin of $S=3$ and a total orbital angular momentum of $L=3$ according to Hund's rules. 
A strong spin-orbit coupling yields a total electronic angular momentum of $J=6$. 
Due to the exceptionally high ligand field induced by the two phthalocyanine ligands (negative axial zero-field splitting) the electron-spin ground state doublet of $J_z=m_J=\pm6$ is well-separated by more than 50~meV from the next higher doublet of $J_z=\pm5$ \cite{Ishikawa2003a,Ishikawa2003b,Ishikawa2004b,Ishikawa2005}.  
This property makes \tbpc~ an exceptional single-molecule magnet \cite{Gatteschi2006} that behaves like an Ising spin at temperatures up to $\approx100~K$ \cite{Ishikawa2003b}. 
Since each phthalocyanine ligand carries a nominal charge of $-2$, the charge state of \tbpc\, molecules is singly negative in solution and bulk phase \cite{Ishikawa2003b,Ishikawa2005}, however, also neutral when adsorbed on a Au(111) surface \cite{Katoh2009,Komeda2011}.

In addition to the electronic spin, the Tb$^{3+}$ ion carries a nuclear spin of $I=3/2$.
A large hyperfine interaction (dipole and quadrupole) between the nuclear spin and the electronic angular momentum causes a fourfold splitting of each electronic spin level. 
The nuclear quadrupole term results in a non-equidistant spacing of the electronic hyperfine levels as illustrated in the Zeeman diagram of Fig.~\ref{fig:tb}b, showing the lowest $J_\mathrm{z}=\pm6$ sub-states of the $J=\pm6$ ground-state doublet of the Tb$^{3+}$ ion in [\tbpc]$^0$. 
The labels on the right side mark the eight different hyperfine levels with $J_\mathrm{z}=\pm6$ and nuclear spin $I=\pm1/2$ and $\pm3/2$, respectively. 
The yellow, green and blue arrows mark ``purely nuclear'' hyperfine transitions, labeled~1--3, with $\Delta I_z=\pm1$ and the electron angular momentum being conserved ($\Delta J_\mathrm{z}=0$). 
The transitions 1--3 depend only weakly on the magnetic field (nuclear Zeeman effect) 
due to the large proton-to-electron mass ratio of $\approx 1836$. 
In comparison, the red arrow marks a ``purely electronic'' hyperfine transition with $\Delta J_\mathrm{z}=\pm12$ and $\Delta I_z=0$, labeled~4, which is considerably Zeeman shifted between the two experimental field values of $B=2.5\pm0.5$ and $16.1\pm0.5$~mT marked by dotted lines. 
In the following, we demonstrate that the different transitions between electronic hyperfine levels of Tb$^{3+}$ shown in Fig.~\ref{fig:tb}b can be resonantly excited and detected by rf-STM.

In order to establish well-defined experimental conditions, throughout the present work, we have studied single neutral [\tbpc]$^0$ molecules on Au(111), which are well-known to adsorb non-dissociatively on Au(111) with the phthalocyanine planes aligned parallel to the substrate surface \cite{Katoh2009}. 
Furthermore, neutral [\tbpc]$^0$ molecules exhibit two electronic spin systems: (i) the Tb$^{3+}$ ion with $J=6$ and (ii) an open-shell $\pi$ system on the phthalocyanine ligands with an electronic spin of $S=1/2$. 
Komeda et al. \cite{Komeda2011} have demonstrated that neutral [\tbpc]$^0$ molecules on Au(111) are clearly identified by a characteristic Kondo signature observed by dc tunneling spectroscopy (Fig.~\ref{fig:tb}c) -- as compared to [\tbpc]$^-$ anions. 
Neutral [\tbpc]$^0$ molecules adopt an achiral geometric configuration \cite{Ishikawa2003b,Komeda2011,Fu2012} characterized by a 45$^\circ$ twist between the lower and upper phthalocyanine ligand (Fig.~\ref{fig:tb}a), showing a characteristic STM topographic appearance \cite{Komeda2011,Fu2012} with a symmetric 8-lobe structure (Fig.~\ref{fig:tb}d). 

The experimental setup is based on a modified commercial STM instrument (Createc) described earlier by our group for the successful detection of rf tunneling current signals in the 100-MHz regime \cite{Mullegger2014a,Mullegger2014b}. 
The electronic hyperfine transitions of \tbpc\, shown in Fig.~\ref{fig:tb}b correspond to transition frequencies, $f=E/h$, between about 100~MHz and 15~GHz. 
Compared to our previous experiments reported in Refs.~\cite{Mullegger2014a,Mullegger2014b}, in the present work, we increased the experimental bandwidth to $4.2$~GHz utilizing improved rf elements. 
By operating our rf-STM at two different static magnetic field values at the sample (marked by dotted lines in Fig.~\ref{fig:tb}b), we have been able to resonantly excite as much as 9 different hyperfine transitions of Tb$^{3+}$ within our experimental bandwidth, including purely nuclear, purely electronic as well as mixed nuclear/electronic spin transitions (see below). 

\begin{figure}
\includegraphics[width=8.6cm]{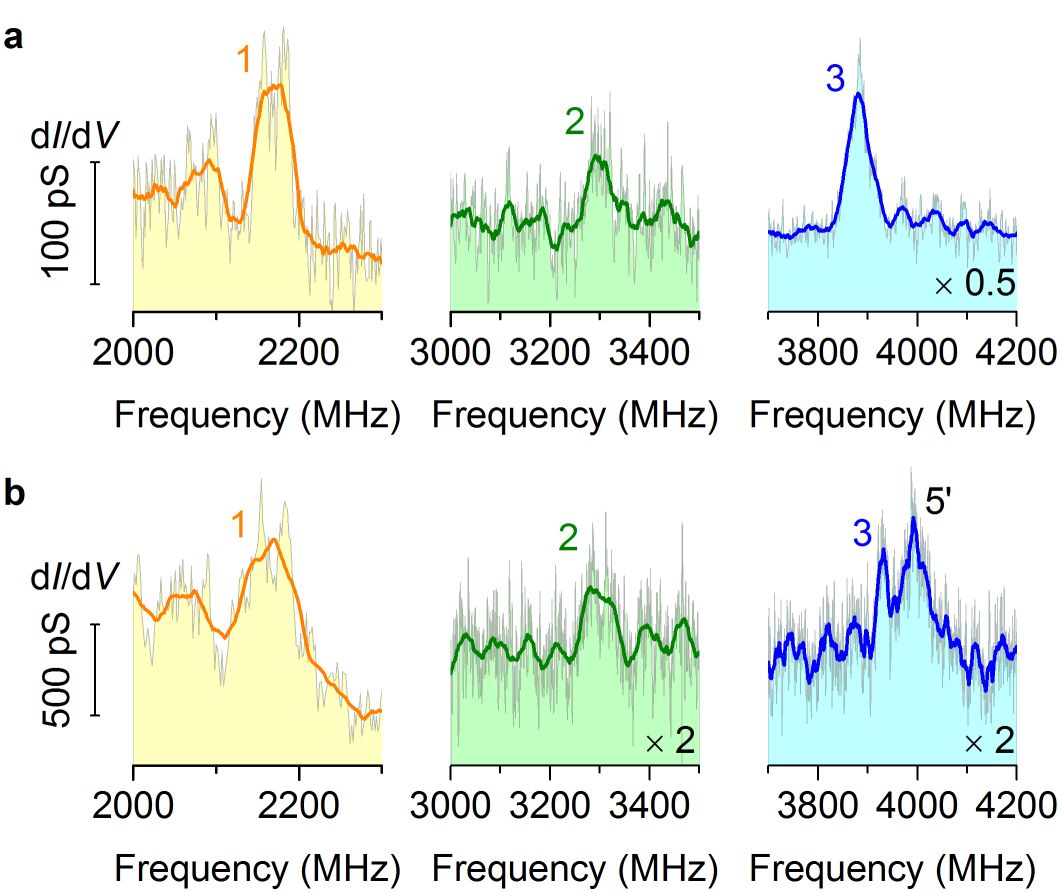}
\caption{\label{fig:nuc} Single-molecule resonance spectra by rf-STM. Conductance (\didv) resonance peaks obtained by rf-STM over single [\tbpc]$^0$ molecules adsorbed on Au(111) at 5~K and a static magnetic field of (a) 2.5~mT and (b) 16.1~mT applied perpendicular to the sample surface; the STM tip was placed over the ligand ($+0.9$~V sample voltage, 10~MHz/s rf sweep rate); solid line indicates averaged adjacent data points in each single spectrum; labels 1--3 and 5' relate to selected spin transitions marked by arrows in Fig.~\ref{fig:tb}b.} 
\end{figure}

For resonant spin excitation, the STM tip is placed at a fixed position over single \tbpc\, molecules at a constant tunneling current between 0.5 and 2~nA. 
The sample voltage was set at $+0.9$~V, facilitating electron tunneling from the STM tip into an unoccupied molecular orbital of the phthalocyanine ligand of \tbpc/Au(111) \cite{Komeda2011}. 
The $\pi$-type electronic structure of phthalocyanine permits coupling between the Tb$^{3+}$ ion and the environment \cite{Lodi2011,Komeda2011}.
The phthalocyanine provides a transport channel, where the conductance of the ligand is coupled to the spin state of the Tb$^{3+}$ ion as shown in recent transport experiments on single \tbpc\, molecules in a molecular spin transistor \cite{Vincent2012} as well as in a supramolecular spin-valve geometry \cite{Urdampilleta2013}. 
In our rf-STM experiments, the dc tunneling conductance, \didv, is measured while simultaneously modulating the sample bias at a variable radio frequency. 
The conductance of the phthalocyanine ligand is affected by the excitation of hyperfine-split electronic states of Tb$^{3+}$ by rf tunneling electrons  -- a process not restricted by the selection rules of photon-induced magnetic dipole transitions. 

\begin{figure}
\includegraphics[width=6.5cm]{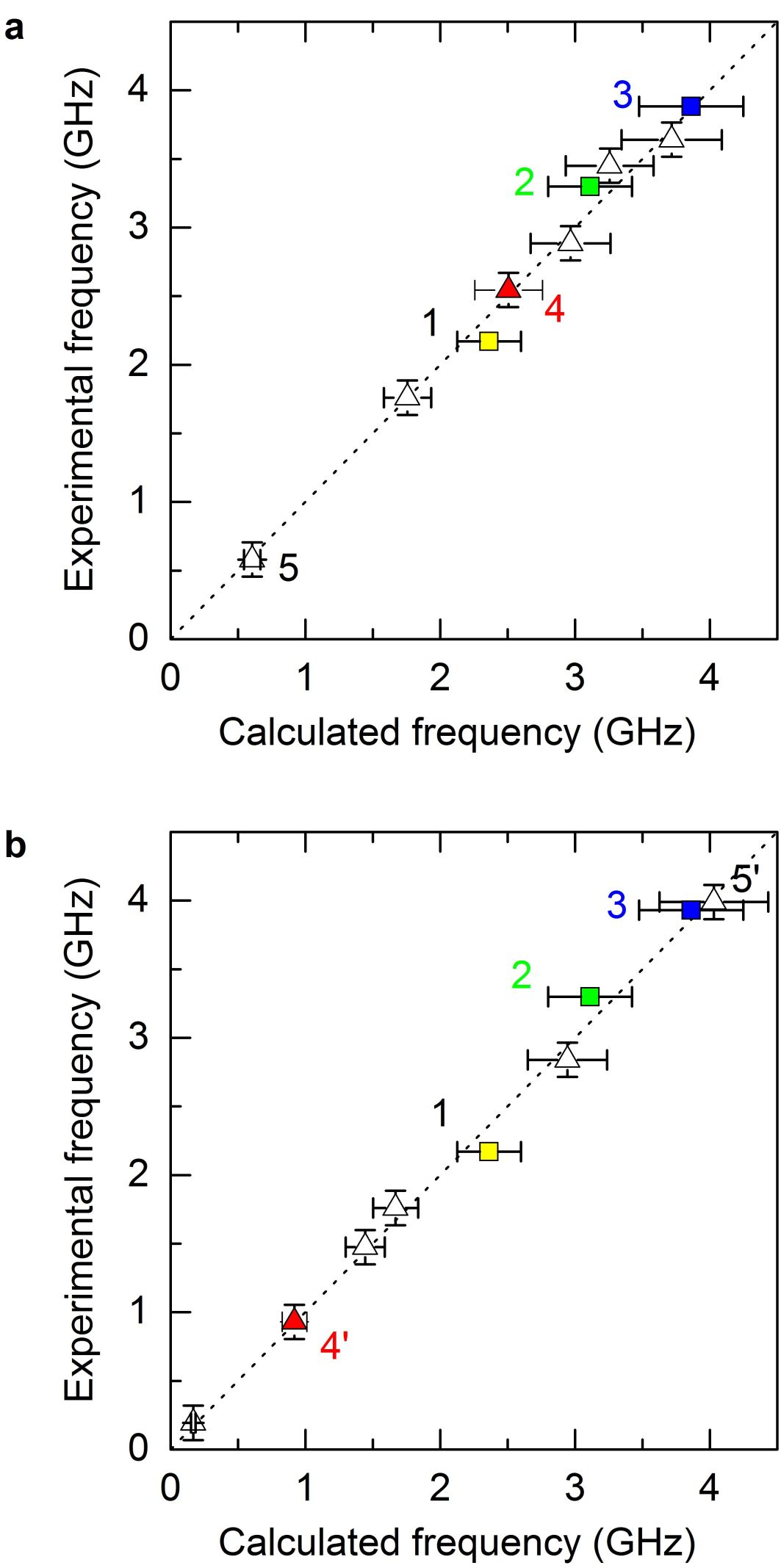}
\caption{\label{fig:exptheor} Experimental and calculated spin transition frequencies of [\tbpc]$^0$. Comparison of the experimental frequency values obtained by rf-STM on single [\tbpc]$^0$ molecules on Au(111) with the calculated hyperfine transition frequencies of \tbpc~ at a static magnetic field of (a) 2.5~mT and (b) 16.1~mT perpendicular to the sample surface; boxes ($\square$) mark purely nuclear transitions; triangles ($\triangle$) mark transitions with electronic spin flips; the dotted line represents ideal agreement as guide to the eye; selected transitions marked by arrows in Fig.~\ref{fig:tb}b are labeled 1--5, 4', and 5'; error bars are $\pm10$\% (horizontal), $\pm0.05$~GHz (vertical, $\square$), and $\pm0.125$~GHz (vertical, $\triangle$), see text.} 
\end{figure}

Intriguingly, with the STM tip over single [\tbpc]$^0$ molecules we observe distinct peaks (increased \didv) at certain rf values, compared to a constant signal over bare Au(111). 
Figure~\ref{fig:nuc}a shows the \didv\, signal obtained over single \tbpc\, molecules while simultaneously ramping the external rf modulation over different frequency ranges 
in the presence of a static magnetic field of $2.5\pm0.5$~mT at the sample. 
The experimental peak positions are in very good agreement with the calculated frequency values of the ``purely nuclear'' hyperfine transitions \mbox{1--3} in Fig.~\ref{fig:tb}b within the measurement error (discussed below).  
The electronic hyperfine levels of Tb$^{3+}$ in [\tbpc]$^0$ (Fig.~\ref{fig:tb}b) are obtained by numerical diagonalization of the spin Hamiltonian analogous to the calculations by Ishikawa et al. \cite{Ishikawa2003b,Ishikawa2005} on negatively charged [\tbpc]$^-$. 
Increasing the static magnetic field to $16.1\pm0.5$~mT leaves the peak positions almost unchanged (Fig.~\ref{fig:nuc}b), which is consistent with the expected weak B-field dependence of the transitions \mbox{1--3}. 
Notice, that transition~3 corresponds to the left peak of Fig.~\ref{fig:exptheor}b,\,right; peak 5' is a mixed electronic-nuclear transition subject to a strong electronic Zeeman shift (compare Fig.~\ref{fig:tb}b). 
Both peaks 3 and 5' in Fig.~\ref{fig:nuc}b lie closely adjacent, but are still discernible as separate peaks; thereof, we derive an approximate spectroscopic resolution of our \mbox{rf-STM} method of $\frac{f}{\Delta f} \approx 70$, which is at least one order of magnitude higher compared to conventional \mbox{dc-STM} spin spectroscopy \cite{Heinrich2004,Fu2009}. 
Repeating the experiment with different STM tips and different \tbpc\, molecules adsorbed over different sites of the Au(111) lattice reproduces the \didv\, peaks within an experimental uncertainty of $\pm50$~MHz. 

Indeed, we observe \didv\, peaks that lie very close to the predicted frequencies of all the possible hyperfine transitions of the lowest $J_z=\pm6$ substates of Tb$^{3+}$ in \tbpc\, that lie within our experimental bandwidth. 
We performed measurements at two different static magnetic fields of $2.5\pm0.5$ and $16.1\pm0.5$~mT at the sample. 
The observed transitions include purely nuclear, purely electronic as well as mixed nuclear-electronic hyperfine transitions. 
Figure~\ref{fig:exptheor} juxtaposes the experimental frequencies obtained over single \tbpc\, molecules on Au(111) with the calculated hyperfine transition frequencies of Tb$^{3+}$ for a static magnetic field of 2.5~mT (Fig.~\ref{fig:exptheor}a) and 16.1~mT (Fig.~\ref{fig:exptheor}b). 
Within the error bars of experiment and calculation, all data points lie on the dotted line (guide to the eye), thus revealing a stunningly good agreement for all observed frequencies with the calculated ones. 
The above results strongly evidence that the frequency values of the resonant \didv\, peaks observed by rf-STM, indeed, relate to the transition frequencies of different hyperfine levels of the Tb$^{3+}$ ion in \tbpc\, adsorbed on Au(111). 
Compared to dc-STM based spin spectroscopy performed \cite{Heinrich2004,Fu2009}, the magnetic excitation energies of only a few micro-electronvolts, detected by our rf-STM, are three orders of magnitudes lower than the mean thermal energy (few milli-electronvolt) of the sample at 5~K. 
Whenever a new method under test, such as our rf-STM, is capable of correctly and unambiguously detecting the manifold of different transition frequencies simultaneously, this can be taken as strong evidence confirming proper functioning. 
In comparison, no signatures of the magnetic moment of the Tb$^{3+}$ ion have been detected by conventional dc-STM tunneling spectroscopy \cite{Komeda2011}. 

In Fig.~\ref{fig:exptheor} the experimental error bars (vertical) of $\pm0.125$~GHz for hyperfine transitions with electronic spin flips $\Delta J_z\neq0$, plotted as triangles ($\triangle$), are dominated by the uncertainty of the static magnetic field of our experimental setup ($\pm0.5$~mT). 
For all transitions, the different adsorption sites of different \tbpc\, molecules investigated by rf-STM cause a measurement uncertainty of about $\pm0.05$~GHz; this is the dominant contribution to the experimental error bar of the ``purely nuclear'' transitions plotted as boxes ($\square$). 
In addition, we consider the uncertainty of the calculated values (horizontal in Fig.~\ref{fig:exptheor}), which are based on numerical parameters obtained from negatively charged [\tbpc]$^-$ in bulk phase \cite{Ishikawa2005}. 
In comparison, single neutral [\tbpc]$^0$ molecules adsorbed on Au(111), studied in the present work, are expected to exhibit a different ligand field and hyperfine parameters due to the different charge state and molecule-substrate interaction -- as recently pointed out by Urdampilleta et al. \cite{Urdampilleta2013}. 
The error bars of $\pm10$\% represent the approximate deviations of the ligand field and hyperfine parameters of the adsorbed neutral molecule compared to bulk phase. 

In conclusion, our rf-STM results evidence the well-controlled resonant excitation and detection of electronic and nuclear spin transitions with single-spin sensitivity and submolecular spatial resolution combined with fast data acquisition. 
Compared to classical magnetic resonance methods, the presented approach utilizes rf tunneling electrons and is, thus, not restricted by the selection rules of magnetic dipole transitions.  
We successfully combined STM with magnetic resonance principles into a new analysis technique easily adaptable to a variety of nanoscale systems in the fields of nanoscience and -technology, molecular electronics/spintronics, quantum computation/information, radical chemistry, chemical and biological catalysis, medical therapeutics. 
This promises unprecedented possibilities for characterizing and controlling physical properties  at the scale of single atoms and molecules -- including spin excitation and resonance spectroscopy, magnetic writing/reading, chemical analysis, and characterization of oxidation state. 

\section*{Methods}
\tbpc\, molecules were synthesized by the procedure reported in Refs.~\onlinecite{DeCian1985,Gonidec2012} followed by purification in a column chromatograph with silicagel and toluene as an eluent. 
The single-crystal Au(111) substrate was prepared by repeated cycles of Ar$^+$ ion bombardement and annealing at 700~K. 
\tbpc\, was thermally evaporated at ultra-high vacuum conditions from a quartz crucible at 700~K after thorough degassing for $>12$~h at 363~K and $<1$~h at 473~K. 
The static magnetic field perpendicular to the sample was obtained by mounting a SmCo permanent magnet close to the sample and measured \textit{ex situ}. 
STM experiments were performed after \textit{in situ} transfer of the sample, employing electrochemically etched W tips deoxidized by annealing in vacuum. 
Impurity- and tip effects were minimized by multiple tip formings between the experiments, resulting in Au-coated tips. 
The \didv\, signal was obtained from the first-harmonic tunneling current signal detected by lock-in technique with sinusoidal modulation of the sample voltage at 704~Hz and 10--20~mV peak-to-peak. 
Reliable tip performance was established by accurately reproducing the characteristic conductance signature of the Au(111) surface state well-known in the literature \cite{Chen1998}. 
For the rf resonance experiments, a sinusoidal ac tunneling voltage from an rf signal generator is coupled in via a bias tee to the sample (further details see Ref.~\cite{Mullegger2014b}). 
The modulation frequency was swept between 150~MHz and 4~GHz, respectively, at a rate of 10~MHz/s. 
The resulting rf current does not affect the tunnel distance, because the radio frequency is several orders of magnitude larger than, both, the cutoff frequency of the STM feedback loop and the bandwidth of the high-gain current amplifier of the STM. 
The maximum rf power level at the sample was $-20$~dBm.

\section*{Acknowledgments}
\begin{acknowledgments}
We kindly acknowledge financial support of the project I958 by the Austrian Science Fund (FWF). 
\end{acknowledgments}


\begin{thebibliography}{25}%
\makeatletter
\providecommand \@ifxundefined [1]{%
 \@ifx{#1\undefined}
}%
\providecommand \@ifnum [1]{%
 \ifnum #1\expandafter \@firstoftwo
 \else \expandafter \@secondoftwo
 \fi
}%
\providecommand \@ifx [1]{%
 \ifx #1\expandafter \@firstoftwo
 \else \expandafter \@secondoftwo
 \fi
}%
\providecommand \natexlab [1]{#1}%
\providecommand \enquote  [1]{``#1''}%
\providecommand \bibnamefont  [1]{#1}%
\providecommand \bibfnamefont [1]{#1}%
\providecommand \citenamefont [1]{#1}%
\providecommand \href@noop [0]{\@secondoftwo}%
\providecommand \href [0]{\begingroup \@sanitize@url \@href}%
\providecommand \@href[1]{\@@startlink{#1}\@@href}%
\providecommand \@@href[1]{\endgroup#1\@@endlink}%
\providecommand \@sanitize@url [0]{\catcode `\\12\catcode `\$12\catcode
  `\&12\catcode `\#12\catcode `\^12\catcode `\_12\catcode `\%12\relax}%
\providecommand \@@startlink[1]{}%
\providecommand \@@endlink[0]{}%
\providecommand \url  [0]{\begingroup\@sanitize@url \@url }%
\providecommand \@url [1]{\endgroup\@href {#1}{\urlprefix }}%
\providecommand \urlprefix  [0]{URL }%
\providecommand \Eprint [0]{\href }%
\providecommand \doibase [0]{http://dx.doi.org/}%
\providecommand \selectlanguage [0]{\@gobble}%
\providecommand \bibinfo  [0]{\@secondoftwo}%
\providecommand \bibfield  [0]{\@secondoftwo}%
\providecommand \translation [1]{[#1]}%
\providecommand \BibitemOpen [0]{}%
\providecommand \bibitemStop [0]{}%
\providecommand \bibitemNoStop [0]{.\EOS\space}%
\providecommand \EOS [0]{\spacefactor3000\relax}%
\providecommand \BibitemShut  [1]{\csname bibitem#1\endcsname}%
\let\auto@bib@innerbib\@empty
\bibitem [{\citenamefont {Chen}(2008)}]{Chen2008}%
  \BibitemOpen
  \bibfield  {author} {\bibinfo {author} {\bibfnamefont {C.~J.}\ \bibnamefont
  {Chen}},\ }\href@noop {} {\emph {\bibinfo {title} {Introduction to Scanning
  Tunneling Microscopy}}},\ \bibinfo {edition} {2nd}\ ed.\ (\bibinfo
  {publisher} {Oxford University Press},\ \bibinfo {year} {2008})\BibitemShut
  {NoStop}%
\bibitem [{\citenamefont {Abragam}\ and\ \citenamefont
  {Bleaney}(1970)}]{Abragam1970}%
  \BibitemOpen
  \bibfield  {author} {\bibinfo {author} {\bibfnamefont {A.}~\bibnamefont
  {Abragam}}\ and\ \bibinfo {author} {\bibfnamefont {B.}~\bibnamefont
  {Bleaney}},\ }\href@noop {} {\emph {\bibinfo {title} {Electron Paramagnetic
  Resonance of Transition Ions}}}\ (\bibinfo  {publisher} {Oxford University
  Press, Oxford},\ \bibinfo {year} {1970})\BibitemShut {NoStop}%
\bibitem [{\citenamefont {Slichter}(1996)}]{Slichter1996}%
  \BibitemOpen
  \bibfield  {author} {\bibinfo {author} {\bibfnamefont {C.~P.}\ \bibnamefont
  {Slichter}},\ }\href@noop {} {\emph {\bibinfo {title} {Principles of Magnetic
  Resonance}}},\ \bibinfo {edition} {3rd}\ ed.\ (\bibinfo  {publisher}
  {Springer-Verlag, Berlin},\ \bibinfo {year} {1996})\BibitemShut {NoStop}%
\bibitem [{\citenamefont {Levitt}(2001)}]{Levitt2001}%
  \BibitemOpen
  \bibfield  {author} {\bibinfo {author} {\bibfnamefont {M.}~\bibnamefont
  {Levitt}},\ }\href@noop {} {\emph {\bibinfo {title} {Spin Dynamics: Basics of
  Nuclear Magnetic Resonance}}}\ (\bibinfo  {publisher} {Wiley},\ \bibinfo
  {year} {2001})\BibitemShut {NoStop}%
\bibitem [{\citenamefont {Balashov}\ \emph {et~al.}(2006)\citenamefont
  {Balashov}, \citenamefont {Takacs}, \citenamefont {Wulfhekel},\ and\
  \citenamefont {Kirschner}}]{Balashov2006}%
  \BibitemOpen
  \bibfield  {author} {\bibinfo {author} {\bibfnamefont {T.}~\bibnamefont
  {Balashov}}, \bibinfo {author} {\bibfnamefont {A.~F.}\ \bibnamefont
  {Takacs}}, \bibinfo {author} {\bibfnamefont {W.}~\bibnamefont {Wulfhekel}}, \
  and\ \bibinfo {author} {\bibfnamefont {J.}~\bibnamefont {Kirschner}},\
  }\href@noop {} {\bibfield  {journal} {\bibinfo  {journal} {Phys. Rev. Lett.}\
  }\textbf {\bibinfo {volume} {97}},\ \bibinfo {pages} {187201} (\bibinfo
  {year} {2006})}\BibitemShut {NoStop}%
\bibitem [{\citenamefont {Heinrich}\ \emph {et~al.}(2004)\citenamefont
  {Heinrich}, \citenamefont {Gupta}, \citenamefont {Lutz},\ and\ \citenamefont
  {Eigler}}]{Heinrich2004}%
  \BibitemOpen
  \bibfield  {author} {\bibinfo {author} {\bibfnamefont {A.~J.}\ \bibnamefont
  {Heinrich}}, \bibinfo {author} {\bibfnamefont {J.~A.}\ \bibnamefont {Gupta}},
  \bibinfo {author} {\bibfnamefont {C.~P.}\ \bibnamefont {Lutz}}, \ and\
  \bibinfo {author} {\bibfnamefont {D.~M.}\ \bibnamefont {Eigler}},\
  }\href@noop {} {\bibfield  {journal} {\bibinfo  {journal} {Science}\ }\textbf
  {\bibinfo {volume} {306}},\ \bibinfo {pages} {466} (\bibinfo {year}
  {2004})}\BibitemShut {NoStop}%
\bibitem [{\citenamefont {Loth}\ \emph {et~al.}(2010)\citenamefont {Loth},
  \citenamefont {Etzkorn}, \citenamefont {Lutz}, \citenamefont {Eigler},\ and\
  \citenamefont {Heinrich}}]{Loth2010}%
  \BibitemOpen
  \bibfield  {author} {\bibinfo {author} {\bibfnamefont {S.}~\bibnamefont
  {Loth}}, \bibinfo {author} {\bibfnamefont {M.}~\bibnamefont {Etzkorn}},
  \bibinfo {author} {\bibfnamefont {C.~P.}\ \bibnamefont {Lutz}}, \bibinfo
  {author} {\bibfnamefont {D.~M.}\ \bibnamefont {Eigler}}, \ and\ \bibinfo
  {author} {\bibfnamefont {A.~J.}\ \bibnamefont {Heinrich}},\ }\href@noop {}
  {\bibfield  {journal} {\bibinfo  {journal} {Science}\ }\textbf {\bibinfo
  {volume} {329}},\ \bibinfo {pages} {1628} (\bibinfo {year}
  {2010})}\BibitemShut {NoStop}%
\bibitem [{\citenamefont {Komeda}\ \emph {et~al.}(2011)\citenamefont {Komeda},
  \citenamefont {Isshiki}, \citenamefont {Liu}, \citenamefont {Zhang},
  \citenamefont {Lorente}, \citenamefont {Kato}, \citenamefont {Breedlove},\
  and\ \citenamefont {Yamashita}}]{Komeda2011}%
  \BibitemOpen
  \bibfield  {author} {\bibinfo {author} {\bibfnamefont {T.}~\bibnamefont
  {Komeda}}, \bibinfo {author} {\bibfnamefont {H.}~\bibnamefont {Isshiki}},
  \bibinfo {author} {\bibfnamefont {J.}~\bibnamefont {Liu}}, \bibinfo {author}
  {\bibfnamefont {Y.~F.}\ \bibnamefont {Zhang}}, \bibinfo {author}
  {\bibfnamefont {N.}~\bibnamefont {Lorente}}, \bibinfo {author} {\bibfnamefont
  {K.}~\bibnamefont {Kato}}, \bibinfo {author} {\bibfnamefont {B.~K.}\
  \bibnamefont {Breedlove}}, \ and\ \bibinfo {author} {\bibfnamefont
  {M.}~\bibnamefont {Yamashita}},\ }\href {\doibase 10.1038/ncomms1210}
  {\bibfield  {journal} {\bibinfo  {journal} {Nature Commun.}\ }\textbf
  {\bibinfo {volume} {2}},\ \bibinfo {pages} {217} (\bibinfo {year}
  {2011})}\BibitemShut {NoStop}%
\bibitem [{\citenamefont {Fang}\ \emph {et~al.}(2011)\citenamefont {Fang},
  \citenamefont {Kurebayashi}, \citenamefont {Wunderlich}, \citenamefont
  {Vyborny}, \citenamefont {Zarbo}, \citenamefont {Campion}, \citenamefont
  {Casiraghi}, \citenamefont {Gallagher}, \citenamefont {Jungwirth},\ and\
  \citenamefont {Ferguson}}]{Fang2011}%
  \BibitemOpen
  \bibfield  {author} {\bibinfo {author} {\bibfnamefont {D.}~\bibnamefont
  {Fang}}, \bibinfo {author} {\bibfnamefont {H.}~\bibnamefont {Kurebayashi}},
  \bibinfo {author} {\bibfnamefont {J.}~\bibnamefont {Wunderlich}}, \bibinfo
  {author} {\bibfnamefont {K.}~\bibnamefont {Vyborny}}, \bibinfo {author}
  {\bibfnamefont {L.~P.}\ \bibnamefont {Zarbo}}, \bibinfo {author}
  {\bibfnamefont {R.~P.}\ \bibnamefont {Campion}}, \bibinfo {author}
  {\bibfnamefont {A.}~\bibnamefont {Casiraghi}}, \bibinfo {author}
  {\bibfnamefont {B.~L.}\ \bibnamefont {Gallagher}}, \bibinfo {author}
  {\bibfnamefont {T.}~\bibnamefont {Jungwirth}}, \ and\ \bibinfo {author}
  {\bibfnamefont {A.~J.}\ \bibnamefont {Ferguson}},\ }\href@noop {} {\bibfield
  {journal} {\bibinfo  {journal} {Nat. Nano.}\ }\textbf {\bibinfo {volume}
  {6}},\ \bibinfo {pages} {413} (\bibinfo {year} {2011})}\BibitemShut {NoStop}%
\bibitem [{\citenamefont {Ishikawa}\ \emph
  {et~al.}(2003{\natexlab{a}})\citenamefont {Ishikawa}, \citenamefont {Sugita},
  \citenamefont {Ishikawa}, \citenamefont {Koshihara},\ and\ \citenamefont
  {Kaizu}}]{Ishikawa2003a}%
  \BibitemOpen
  \bibfield  {author} {\bibinfo {author} {\bibfnamefont {N.}~\bibnamefont
  {Ishikawa}}, \bibinfo {author} {\bibfnamefont {M.}~\bibnamefont {Sugita}},
  \bibinfo {author} {\bibfnamefont {T.}~\bibnamefont {Ishikawa}}, \bibinfo
  {author} {\bibfnamefont {S.}~\bibnamefont {Koshihara}}, \ and\ \bibinfo
  {author} {\bibfnamefont {Y.}~\bibnamefont {Kaizu}},\ }\href@noop {}
  {\bibfield  {journal} {\bibinfo  {journal} {J. Am. Chem. Soc.}\ }\textbf
  {\bibinfo {volume} {125}},\ \bibinfo {pages} {8694} (\bibinfo {year}
  {2003}{\natexlab{a}})}\BibitemShut {NoStop}%
\bibitem [{\citenamefont {Ishikawa}\ \emph
  {et~al.}(2003{\natexlab{b}})\citenamefont {Ishikawa}, \citenamefont {Sugita},
  \citenamefont {Okubo}, \citenamefont {Tanaka}, \citenamefont {Iino},\ and\
  \citenamefont {Kaizu}}]{Ishikawa2003b}%
  \BibitemOpen
  \bibfield  {author} {\bibinfo {author} {\bibfnamefont {N.}~\bibnamefont
  {Ishikawa}}, \bibinfo {author} {\bibfnamefont {M.}~\bibnamefont {Sugita}},
  \bibinfo {author} {\bibfnamefont {T.}~\bibnamefont {Okubo}}, \bibinfo
  {author} {\bibfnamefont {N.}~\bibnamefont {Tanaka}}, \bibinfo {author}
  {\bibfnamefont {T.}~\bibnamefont {Iino}}, \ and\ \bibinfo {author}
  {\bibfnamefont {Y.}~\bibnamefont {Kaizu}},\ }\href@noop {} {\bibfield
  {journal} {\bibinfo  {journal} {Inorg. Chem.}\ }\textbf {\bibinfo {volume}
  {42}},\ \bibinfo {pages} {2440} (\bibinfo {year}
  {2003}{\natexlab{b}})}\BibitemShut {NoStop}%
\bibitem [{\citenamefont {Ishikawa}\ \emph {et~al.}(2004)\citenamefont
  {Ishikawa}, \citenamefont {Sugita}, \citenamefont {Ishikawa}, \citenamefont
  {Koshihara},\ and\ \citenamefont {Kaizu}}]{Ishikawa2004b}%
  \BibitemOpen
  \bibfield  {author} {\bibinfo {author} {\bibfnamefont {N.}~\bibnamefont
  {Ishikawa}}, \bibinfo {author} {\bibfnamefont {M.}~\bibnamefont {Sugita}},
  \bibinfo {author} {\bibfnamefont {T.}~\bibnamefont {Ishikawa}}, \bibinfo
  {author} {\bibfnamefont {S.}~\bibnamefont {Koshihara}}, \ and\ \bibinfo
  {author} {\bibfnamefont {Y.}~\bibnamefont {Kaizu}},\ }\bibfield  {booktitle}
  {\emph {\bibinfo {booktitle} {The Journal of Physical Chemistry B}},\
  }\href@noop {} {\bibfield  {journal} {\bibinfo  {journal} {J. Phys. Chem. B}\
  }\textbf {\bibinfo {volume} {108}},\ \bibinfo {pages} {11265} (\bibinfo
  {year} {2004})}\BibitemShut {NoStop}%
\bibitem [{\citenamefont {Ishikawa}\ \emph {et~al.}(2005)\citenamefont
  {Ishikawa}, \citenamefont {Sugita},\ and\ \citenamefont
  {Wernsdorfer}}]{Ishikawa2005}%
  \BibitemOpen
  \bibfield  {author} {\bibinfo {author} {\bibfnamefont {N.}~\bibnamefont
  {Ishikawa}}, \bibinfo {author} {\bibfnamefont {M.}~\bibnamefont {Sugita}}, \
  and\ \bibinfo {author} {\bibfnamefont {W.}~\bibnamefont {Wernsdorfer}},\
  }\href@noop {} {\bibfield  {journal} {\bibinfo  {journal} {Angew. Chem. Int.
  Edit.}\ }\textbf {\bibinfo {volume} {44}},\ \bibinfo {pages} {2931} (\bibinfo
  {year} {2005})}\BibitemShut {NoStop}%
\bibitem [{\citenamefont {Gatteschi}\ \emph {et~al.}(2006)\citenamefont
  {Gatteschi}, \citenamefont {Sessoli},\ and\ \citenamefont
  {Villain}}]{Gatteschi2006}%
  \BibitemOpen
  \bibfield  {author} {\bibinfo {author} {\bibfnamefont {D.}~\bibnamefont
  {Gatteschi}}, \bibinfo {author} {\bibfnamefont {R.}~\bibnamefont {Sessoli}},
  \ and\ \bibinfo {author} {\bibfnamefont {J.}~\bibnamefont {Villain}},\
  }\href@noop {} {\emph {\bibinfo {title} {Molecular Nanomagnets}}}\ (\bibinfo
  {publisher} {Oxford University Press},\ \bibinfo {year} {2006})\BibitemShut
  {NoStop}%
\bibitem [{\citenamefont {Katoh}\ \emph {et~al.}(2009)\citenamefont {Katoh},
  \citenamefont {Yoshida}, \citenamefont {Yamashita}, \citenamefont {Miyasaka},
  \citenamefont {Breedlove}, \citenamefont {Kajiwara}, \citenamefont
  {Takaishi}, \citenamefont {Ishikawa}, \citenamefont {Isshiki}, \citenamefont
  {Zhang}, \citenamefont {Komeda}, \citenamefont {Yamagishi},\ and\
  \citenamefont {Takeya}}]{Katoh2009}%
  \BibitemOpen
  \bibfield  {author} {\bibinfo {author} {\bibfnamefont {K.}~\bibnamefont
  {Katoh}}, \bibinfo {author} {\bibfnamefont {Y.}~\bibnamefont {Yoshida}},
  \bibinfo {author} {\bibfnamefont {M.}~\bibnamefont {Yamashita}}, \bibinfo
  {author} {\bibfnamefont {H.}~\bibnamefont {Miyasaka}}, \bibinfo {author}
  {\bibfnamefont {B.~K.}\ \bibnamefont {Breedlove}}, \bibinfo {author}
  {\bibfnamefont {T.}~\bibnamefont {Kajiwara}}, \bibinfo {author}
  {\bibfnamefont {S.}~\bibnamefont {Takaishi}}, \bibinfo {author}
  {\bibfnamefont {N.}~\bibnamefont {Ishikawa}}, \bibinfo {author}
  {\bibfnamefont {H.}~\bibnamefont {Isshiki}}, \bibinfo {author} {\bibfnamefont
  {Y.~F.}\ \bibnamefont {Zhang}}, \bibinfo {author} {\bibfnamefont
  {T.}~\bibnamefont {Komeda}}, \bibinfo {author} {\bibfnamefont
  {M.}~\bibnamefont {Yamagishi}}, \ and\ \bibinfo {author} {\bibfnamefont
  {J.}~\bibnamefont {Takeya}},\ }\bibfield  {booktitle} {\emph {\bibinfo
  {booktitle} {Journal of the American Chemical Society}},\ }\href@noop {}
  {\bibfield  {journal} {\bibinfo  {journal} {J. Am. Chem. Soc.}\ }\textbf
  {\bibinfo {volume} {131}},\ \bibinfo {pages} {9967} (\bibinfo {year}
  {2009})}\BibitemShut {NoStop}%
\bibitem [{\citenamefont {Fu}\ \emph {et~al.}(2012)\citenamefont {Fu},
  \citenamefont {Schw\"{o}bel}, \citenamefont {Hla}, \citenamefont {Dilullo},
  \citenamefont {Hoffmann}, \citenamefont {Klyatskaya}, \citenamefont {Ruben},\
  and\ \citenamefont {Wiesendanger}}]{Fu2012}%
  \BibitemOpen
  \bibfield  {author} {\bibinfo {author} {\bibfnamefont {Y.-S.}\ \bibnamefont
  {Fu}}, \bibinfo {author} {\bibfnamefont {J.}~\bibnamefont {Schw\"{o}bel}},
  \bibinfo {author} {\bibfnamefont {S.-W.}\ \bibnamefont {Hla}}, \bibinfo
  {author} {\bibfnamefont {A.}~\bibnamefont {Dilullo}}, \bibinfo {author}
  {\bibfnamefont {G.}~\bibnamefont {Hoffmann}}, \bibinfo {author}
  {\bibfnamefont {S.}~\bibnamefont {Klyatskaya}}, \bibinfo {author}
  {\bibfnamefont {M.}~\bibnamefont {Ruben}}, \ and\ \bibinfo {author}
  {\bibfnamefont {R.}~\bibnamefont {Wiesendanger}},\ }\bibfield  {booktitle}
  {\emph {\bibinfo {booktitle} {Nano Letters}},\ }\href@noop {} {\bibfield
  {journal} {\bibinfo  {journal} {Nano Lett.}\ }\textbf {\bibinfo {volume}
  {12}},\ \bibinfo {pages} {3931} (\bibinfo {year} {2012})}\BibitemShut
  {NoStop}%
\bibitem [{\citenamefont {M\"{u}llegger}\ \emph
  {et~al.}(2014{\natexlab{a}})\citenamefont {M\"{u}llegger}, \citenamefont
  {Rashidi}, \citenamefont {Mayr}, \citenamefont {Fattinger}, \citenamefont
  {Ney},\ and\ \citenamefont {Koch}}]{Mullegger2014a}%
  \BibitemOpen
  \bibfield  {author} {\bibinfo {author} {\bibfnamefont {S.}~\bibnamefont
  {M\"{u}llegger}}, \bibinfo {author} {\bibfnamefont {M.}~\bibnamefont
  {Rashidi}}, \bibinfo {author} {\bibfnamefont {K.}~\bibnamefont {Mayr}},
  \bibinfo {author} {\bibfnamefont {M.}~\bibnamefont {Fattinger}}, \bibinfo
  {author} {\bibfnamefont {A.}~\bibnamefont {Ney}}, \ and\ \bibinfo {author}
  {\bibfnamefont {R.}~\bibnamefont {Koch}},\ }\href@noop {} {\bibfield
  {journal} {\bibinfo  {journal} {Phys. Rev. Lett.}\ }\textbf {\bibinfo
  {volume} {112}},\ \bibinfo {pages} {117201} (\bibinfo {year}
  {2014}{\natexlab{a}})}\BibitemShut {NoStop}%
\bibitem [{\citenamefont {M\"{u}llegger}\ \emph
  {et~al.}(2014{\natexlab{b}})\citenamefont {M\"{u}llegger}, \citenamefont
  {Das}, \citenamefont {Mayr},\ and\ \citenamefont {Koch}}]{Mullegger2014b}%
  \BibitemOpen
  \bibfield  {author} {\bibinfo {author} {\bibfnamefont {S.}~\bibnamefont
  {M\"{u}llegger}}, \bibinfo {author} {\bibfnamefont {A.~K.}\ \bibnamefont
  {Das}}, \bibinfo {author} {\bibfnamefont {K.}~\bibnamefont {Mayr}}, \ and\
  \bibinfo {author} {\bibfnamefont {R.}~\bibnamefont {Koch}},\ }\href@noop {}
  {\bibfield  {journal} {\bibinfo  {journal} {Nanotechnology}\ }\textbf
  {\bibinfo {volume} {25}},\ \bibinfo {pages} {135705} (\bibinfo {year}
  {2014}{\natexlab{b}})}\BibitemShut {NoStop}%
\bibitem [{\citenamefont {Lodi~Rizzini}\ \emph {et~al.}(2011)\citenamefont
  {Lodi~Rizzini}, \citenamefont {Krull}, \citenamefont {Balashov},
  \citenamefont {Kavich}, \citenamefont {Mugarza}, \citenamefont {Miedema},
  \citenamefont {Thakur}, \citenamefont {Sessi}, \citenamefont {Klyatskaya},
  \citenamefont {Ruben}, \citenamefont {Stepanow},\ and\ \citenamefont
  {Gambardella}}]{Lodi2011}%
  \BibitemOpen
  \bibfield  {author} {\bibinfo {author} {\bibfnamefont {A.}~\bibnamefont
  {Lodi~Rizzini}}, \bibinfo {author} {\bibfnamefont {C.}~\bibnamefont {Krull}},
  \bibinfo {author} {\bibfnamefont {T.}~\bibnamefont {Balashov}}, \bibinfo
  {author} {\bibfnamefont {J.~J.}\ \bibnamefont {Kavich}}, \bibinfo {author}
  {\bibfnamefont {A.}~\bibnamefont {Mugarza}}, \bibinfo {author} {\bibfnamefont
  {P.~S.}\ \bibnamefont {Miedema}}, \bibinfo {author} {\bibfnamefont {P.~K.}\
  \bibnamefont {Thakur}}, \bibinfo {author} {\bibfnamefont {V.}~\bibnamefont
  {Sessi}}, \bibinfo {author} {\bibfnamefont {S.}~\bibnamefont {Klyatskaya}},
  \bibinfo {author} {\bibfnamefont {M.}~\bibnamefont {Ruben}}, \bibinfo
  {author} {\bibfnamefont {S.}~\bibnamefont {Stepanow}}, \ and\ \bibinfo
  {author} {\bibfnamefont {P.}~\bibnamefont {Gambardella}},\ }\href@noop {}
  {\bibfield  {journal} {\bibinfo  {journal} {Phys. Rev. Lett.}\ }\textbf
  {\bibinfo {volume} {107}},\ \bibinfo {pages} {177205} (\bibinfo {year}
  {2011})}\BibitemShut {NoStop}%
\bibitem [{\citenamefont {Vincent}\ \emph {et~al.}(2012)\citenamefont
  {Vincent}, \citenamefont {Klyatskaya}, \citenamefont {Ruben}, \citenamefont
  {Wernsdorfer},\ and\ \citenamefont {Balestro}}]{Vincent2012}%
  \BibitemOpen
  \bibfield  {author} {\bibinfo {author} {\bibfnamefont {R.}~\bibnamefont
  {Vincent}}, \bibinfo {author} {\bibfnamefont {S.}~\bibnamefont {Klyatskaya}},
  \bibinfo {author} {\bibfnamefont {M.}~\bibnamefont {Ruben}}, \bibinfo
  {author} {\bibfnamefont {W.}~\bibnamefont {Wernsdorfer}}, \ and\ \bibinfo
  {author} {\bibfnamefont {F.}~\bibnamefont {Balestro}},\ }\href@noop {}
  {\bibfield  {journal} {\bibinfo  {journal} {Nature}\ }\textbf {\bibinfo
  {volume} {488}},\ \bibinfo {pages} {357} (\bibinfo {year}
  {2012})}\BibitemShut {NoStop}%
\bibitem [{\citenamefont {Urdampilleta}\ \emph {et~al.}(2013)\citenamefont
  {Urdampilleta}, \citenamefont {Klyatskaya}, \citenamefont {Ruben},\ and\
  \citenamefont {Wernsdorfer}}]{Urdampilleta2013}%
  \BibitemOpen
  \bibfield  {author} {\bibinfo {author} {\bibfnamefont {M.}~\bibnamefont
  {Urdampilleta}}, \bibinfo {author} {\bibfnamefont {S.}~\bibnamefont
  {Klyatskaya}}, \bibinfo {author} {\bibfnamefont {M.}~\bibnamefont {Ruben}}, \
  and\ \bibinfo {author} {\bibfnamefont {W.}~\bibnamefont {Wernsdorfer}},\
  }\href@noop {} {\bibfield  {journal} {\bibinfo  {journal} {Phys. Rev. B}\
  }\textbf {\bibinfo {volume} {87}},\ \bibinfo {pages} {195412} (\bibinfo
  {year} {2013})}\BibitemShut {NoStop}%
\bibitem [{\citenamefont {Fu}\ \emph {et~al.}(2009)\citenamefont {Fu},
  \citenamefont {Zhang}, \citenamefont {Ji}, \citenamefont {Chen},
  \citenamefont {Ma}, \citenamefont {Jia},\ and\ \citenamefont {Xue}}]{Fu2009}%
  \BibitemOpen
  \bibfield  {author} {\bibinfo {author} {\bibfnamefont {Y.-S.}\ \bibnamefont
  {Fu}}, \bibinfo {author} {\bibfnamefont {T.}~\bibnamefont {Zhang}}, \bibinfo
  {author} {\bibfnamefont {S.-H.}\ \bibnamefont {Ji}}, \bibinfo {author}
  {\bibfnamefont {X.}~\bibnamefont {Chen}}, \bibinfo {author} {\bibfnamefont
  {X.-C.}\ \bibnamefont {Ma}}, \bibinfo {author} {\bibfnamefont {J.-F.}\
  \bibnamefont {Jia}}, \ and\ \bibinfo {author} {\bibfnamefont {Q.-K.}\
  \bibnamefont {Xue}},\ }\href@noop {} {\bibfield  {journal} {\bibinfo
  {journal} {Phys. Rev. Lett.}\ }\textbf {\bibinfo {volume} {103}},\ \bibinfo
  {pages} {257202} (\bibinfo {year} {2009})}\BibitemShut {NoStop}%
\bibitem [{\citenamefont {De~Cian}\ \emph {et~al.}(1985)\citenamefont
  {De~Cian}, \citenamefont {Moussavi}, \citenamefont {Fischer},\ and\
  \citenamefont {Weiss}}]{DeCian1985}%
  \BibitemOpen
  \bibfield  {author} {\bibinfo {author} {\bibfnamefont {A.}~\bibnamefont
  {De~Cian}}, \bibinfo {author} {\bibfnamefont {M.}~\bibnamefont {Moussavi}},
  \bibinfo {author} {\bibfnamefont {J.}~\bibnamefont {Fischer}}, \ and\
  \bibinfo {author} {\bibfnamefont {R.}~\bibnamefont {Weiss}},\ }\bibfield
  {booktitle} {\emph {\bibinfo {booktitle} {Inorganic Chemistry}},\ }\href@noop
  {} {\bibfield  {journal} {\bibinfo  {journal} {Inorg. Chem.}\ }\textbf
  {\bibinfo {volume} {24}},\ \bibinfo {pages} {3162} (\bibinfo {year}
  {1985})}\BibitemShut {NoStop}%
\bibitem [{\citenamefont {Gonidec}\ \emph {et~al.}(2012)\citenamefont
  {Gonidec}, \citenamefont {Amabilino},\ and\ \citenamefont
  {Veciana}}]{Gonidec2012}%
  \BibitemOpen
  \bibfield  {author} {\bibinfo {author} {\bibfnamefont {M.}~\bibnamefont
  {Gonidec}}, \bibinfo {author} {\bibfnamefont {D.~B.}\ \bibnamefont
  {Amabilino}}, \ and\ \bibinfo {author} {\bibfnamefont {J.}~\bibnamefont
  {Veciana}},\ }\href@noop {} {\bibfield  {journal} {\bibinfo  {journal}
  {Dalton Trans.}\ }\textbf {\bibinfo {volume} {41}},\ \bibinfo {pages} {13632}
  (\bibinfo {year} {2012})}\BibitemShut {NoStop}%
\bibitem [{\citenamefont {Chen}\ \emph {et~al.}(1998)\citenamefont {Chen},
  \citenamefont {Madhavan}, \citenamefont {Jamneala},\ and\ \citenamefont
  {Crommie}}]{Chen1998}%
  \BibitemOpen
  \bibfield  {author} {\bibinfo {author} {\bibfnamefont {W.}~\bibnamefont
  {Chen}}, \bibinfo {author} {\bibfnamefont {V.}~\bibnamefont {Madhavan}},
  \bibinfo {author} {\bibfnamefont {T.}~\bibnamefont {Jamneala}}, \ and\
  \bibinfo {author} {\bibfnamefont {M.}~\bibnamefont {Crommie}},\ }\href@noop
  {} {\bibfield  {journal} {\bibinfo  {journal} {Phys. Rev. Lett.}\ }\textbf
  {\bibinfo {volume} {80}},\ \bibinfo {pages} {1469 } (\bibinfo {year}
  {1998})}\BibitemShut {NoStop}%
\end{thebibliography}
\end{document}